# ON ENTROPY WIND IN SUPERFLUID HELIUM

N. I. Pushkina

Moscow State University, Scientific Research Computation Centre, Vorob'yovy Gory, Moscow, 119992  Russia
E-mail: N.Pushkina@mererand.com

**Abstract**

Generation of a quasi-stationary flow of the superfluid helium normal part in the presence of intense first- and second-sound waves is studied. Relevant equations are obtained. The contribution to the process of energy dissipation at the shock front layer and of fluid viscosity is analyzed in detail for the case of a second-sound wave. An estimate concerning possible experimental observation of the process is made.

## 1. Introduction

It is well known that intense sound waves can induce quasi-stationary movements of fluid and gas media. In nonlinear acoustics such movements are called acoustic streams or sound wind. In classical hydrodynamics various types of acoustic flows have been studied (see e.g. [1, 2]). An intense sound wave creates a force that generates a stationary flow due to both nonlinearity of a medium and sound absorption. Sound absorption is essential since through this acoustic wave energy is transferred from the sound field to the flow.

In quantum fluids a number of nonlinear wave phenomena proper also to nonlinear acoustics of classical media have been studied (see Review [3]). But as to a stationary flow induced by intense sound waves this has not yet been investigated in superfluid helium. Concerning stationary flows an essential difference between classical media and quantum liquids is that in superfluids not the whole liquid but only the normal part is carried by the flow. In this sense one may speak of "entropy wind". Besides, one should take into account that the first- and second-sound attenuation in helium is rather small which leads to the formation of a shock wave with a narrow front. In this case not viscosity, but energy dissipation in the shock front layer is of the main importance in the generation of a stationary flow. In the present paper generation of a stationary entropy flow induced by intense first- and second-sound waves in superfluid He $^4$  is studied.

## 2. Theory of the process

For studying stationary flows produced by propagating waves it is natural to average the whole movement over a period of time greater than the wave period, but less than the characteristic time scale of the quasi-stationary flow. To describe a slow (compared with the wave velocity) stationary flow an approximation of incompressible fluid is valid not only as to the whole density ($\rho_0$ = const), but also as to the normal ($\rho_{n0}$ = const) and superfluid ($\rho_{s0}$ = const) densities; the order of smallness of dissipative terms in the entropy equation makes it possible to consider $\sigma$ = const, where $\sigma$ is the entropy per unit mass (see [4]). Since $\rho_0$, $\rho_{no}$, $\rho_{s0}$, $\sigma$ are constants the two-fluid hydrodynamic equations yield $\text{div}\,\mathbf{U}_n = \text{div}\,\mathbf{U}_s = 0$, where $\mathbf{U}_n$, $\mathbf{U}_s$ — are the normal and superfluid velocities in the stationary flow.

To find forces produced by first- and second sound waves which induce stationary flow we use the equation for the momentum **j** taking into account dissipative terms

$$\frac{\partial j_i}{\partial t} + \frac{\partial \Pi_{ik}}{\partial r_k} = \frac{\partial}{\partial r_k}\left\{\eta\left(\frac{\partial v_{ni}}{\partial r_k} + \frac{\partial v_{nk}}{\partial r_i} - \frac{2}{3}\delta_{ik}\frac{\partial v_{nl}}{\partial r_l}\right) + \delta_{ik}\xi_1\,\text{div}\,(\mathbf{j} - \rho\mathbf{v}_n) + \delta_{ik}\xi_2\,\text{div}\,\mathbf{v}_n\right\} \quad (1)$$

We put into this equation the densities, pressure and the velocities in the form:

$$\rho = \rho_0 + \rho_a, \quad \rho_n = \rho_{n0} + \rho_{na}, \quad \rho_s = \rho_{s0} + \rho_{sa}, \quad p = p_0 + p_a,$$
$$\mathbf{v}_n = \mathbf{v}_{n0} + \mathbf{v}_{na}, \quad \mathbf{v}_s = \mathbf{v}_{s0} + \mathbf{v}_{sa}, \quad (2)$$

where the quantities with the index "*a*", averaged over a period of time greater than the first-sound or the temperature wave period, are equal to zero and refer to waves and those with the index "0" characterize the flow. Putting (2) in the continuity equation

$$\frac{\partial \rho}{\partial t} + \text{div}\,\mathbf{j} = 0,$$

averaging it over time and taking into account nonlinearity, we obtain the following equality

$$\rho_{n0}\,\text{div}\left(\mathbf{v}_{n0} + \frac{\overline{\rho_{na}\mathbf{v}_{na}}}{\rho_{n0}}\right) + \rho_{s0}\,\text{div}\left(\mathbf{v}_{s0} + \frac{\overline{\rho_{sa}\mathbf{v}_{sa}}}{\rho_{s0}}\right) = 0.$$

From this one can draw a conclusion that due to nonlinearity the flow velocities $\mathbf{U}_n$ and $\mathbf{U}_s$ equal actually not $\mathbf{v}_{n0}$ and $\mathbf{v}_{s0}$, but the following expressions

$$\mathbf{U}_n = \mathbf{v}_{n0} + \frac{\overline{\rho_{na}\mathbf{v}_{na}}}{\rho_{n0}}, \qquad \mathbf{U}_s = \mathbf{v}_{s0} + \frac{\overline{\rho_{sa}\mathbf{v}_{sa}}}{\rho_{s0}}. \tag{3}$$

Putting the relations (2), (3) into Eq. (1) and averaging it over time we obtain the following equation

$$\frac{\partial \mathbf{U}_n}{\partial t} + (\mathbf{U}_n \nabla)\mathbf{U}_n - \frac{\eta}{\rho_{n0}}\Delta \mathbf{U}_n + \frac{1}{\rho_{n0}}\nabla\left(\rho_{s0}\frac{U_s^2}{2} + \rho_{s0}\frac{\partial \varphi}{\partial t} + p_{s0}\right) = -\frac{1}{\rho_{n0}}\nabla p_{n0} + \mathbf{F}_{1,2}, \tag{4}$$

where $\mathbf{F}_{1,2}$ refer to the first- and second-sound waves:

$$\mathbf{F}_1 = -\frac{\rho_0}{\rho_{n0}}\left[\overline{(\mathbf{v}_{na}\nabla)\mathbf{v}_{na}} + \overline{\mathbf{v}_{na}\operatorname{div}\mathbf{v}_{na}}\right] - \frac{\eta}{\rho_{n0}^2}\Delta\left(\overline{\rho_{na}\mathbf{v}_{na}}\right) -$$

$$\frac{1}{\rho_{n0}^2}\left(\frac{\eta}{3} + \xi_2 - \rho_0\xi_1\right)\operatorname{grad}\operatorname{div}\left(\overline{\rho_{na}\mathbf{v}_{na}}\right) - \frac{1}{\rho_{n0}}\xi_1\operatorname{grad}\operatorname{div}\left(\overline{\rho_a\mathbf{v}_{na}}\right);$$

$$\mathbf{F}_2 = -\frac{\rho_0}{\rho_{s0}}\left[\overline{(\mathbf{v}_{na}\nabla)\mathbf{v}_{na}} + \overline{\mathbf{v}_{na}\operatorname{div}\mathbf{v}_{na}}\right] - \frac{\eta}{\rho_{n0}^2}\Delta\left(\overline{\rho_{na}\mathbf{v}_{na}}\right) - \frac{1}{\rho_{n0}^2}\left(\frac{\eta}{3} + \xi_2 - \rho_0\xi_1\right)\operatorname{grad}\operatorname{div}\left(\overline{\rho_{na}\mathbf{v}_{na}}\right);$$

here $\eta$, $\xi_2$, $\xi_1$ are the viscosities.

In Eq. (4) we use as auxiliary quantities the "pressures" of the normal and superfluid parts, $p_{n0}$ and $p_{s0}$ (see [4]) in accordance with the equality $p_0 = p_\infty + p_{n0} + p_{s0}$, where $p_\infty$ is the pressure at the infinity and $p_{s0}$ is defined by the formula valid for an ideal fluid

$$p_{s0} = -\rho_{s0}\frac{\partial \varphi}{\partial t} - \frac{\rho_{s0}U_s^2}{2}, \tag{5}$$

where $\varphi$ is the superfluid potential, $\mathbf{U}_s = \nabla \varphi$. Due to the equality (5) the superfluid velocity $\mathbf{U}_s$ vanishes from Eq. (4). In what follows we omit the index "0".

The quantities $\mathbf{F}_1$ и $\mathbf{F}_2$ can be viewed as the forces acting on the normal unit mass and generating stationary flow. These forces are produced by intense first- and second-sound decaying waves. Wave dissipation is due to the viscosity of a medium and to the energy dissipation in a shock wave front. Note that the extreme smallness of the thermal expansion coefficient in superfluid helium has been taken into account at deriving the expressions for $\mathbf{F}_1$ and $\mathbf{F}_2$.

We record here the one-dimensional equations, all variables are assumed to depend only on the $x$ coordinate and time $t$, and all motion takes place in the $x$ direction. In this case the forces $F_1$ and $F_2$, in terms of the variable $v_{na}$, take the form

$$F_1 = -\frac{\rho}{\rho_n} \frac{\overline{\partial v_{na}^2}}{\partial x} - \frac{\rho}{\rho_n} c_1 \left[ \frac{1}{\rho_n} \frac{\partial \rho_n}{\partial p} \left( \frac{4}{3}\eta + \xi_2 \right) - \left( \frac{\rho}{\rho_n} \frac{\partial \rho_n}{\partial p} + \frac{1}{c_1^2} \right) \xi_1 \right] \frac{\overline{\partial^2 v_{na}^2}}{\partial x^2};$$

$$F_2 = -\frac{\rho}{\rho_s} \frac{\overline{\partial v_{na}^2}}{\partial x} - \frac{c_2}{\sigma \rho_s} \frac{1}{\rho_n} \frac{\partial \rho_n}{\partial T} \left( \frac{4}{3}\eta + \xi_2 - \rho \xi_1 \right) \frac{\overline{\partial^2 v_{na}^2}}{\partial x^2}, \qquad (6)$$

$c_1$ and $c_2$ are the first- and second-sound velocities. $F_1$ is somewhat different from that in classical acoustics, in particular, here there is an extra viscosity coefficient that is usually denoted as $\xi_1$ in the two-fluid hydrodynamics. Below we shall dwell on the case of the second sound only.

To describe nonlinear second-sound wave propagation with dissipation we shall use Burgers equation in the form [3]:

$$\frac{\partial v_{na}}{\partial t} + \left( c_2 + \alpha_2(T) v_n \right) \frac{\partial v_{na}}{\partial x} = \mu \frac{\partial^2 v_{na}}{\partial x^2}, \qquad (7)$$

where $\alpha_2(T) = \frac{\sigma T}{c} \frac{\partial}{\partial T} \ln \left( c_2^3 \frac{\partial \sigma}{\partial T} \right)$, $c$ being the specific heat per unit mass,

$$\mu = \frac{\rho_s}{2\rho\rho_n} \left( \frac{4}{3}\eta + \xi_2 - 2\rho\xi_1 + \rho^2\xi_3 + \frac{\rho_n}{\rho_s} \frac{\kappa}{c} \right), \quad \kappa \text{ is the thermal conductivity.}$$

In Eq. (7) it is convenient to use a reference system moving with the velocity $c_2$: $\tau = t - x/c_2$, $x \to x$. With this we obtain Eq. (7) in the form

$$\frac{\partial v_{na}}{\partial x} - \frac{\alpha_2(T)}{c_2^2} v_{na} \frac{\partial v_{na}}{\partial \tau} = \frac{\mu}{c_2^2} \frac{\partial^2 v_{na}}{\partial \tau^2}. \qquad (8)$$

By solving Eq. (8) one can get an explicit expression for the force $F_2$.

The relative contribution of nonlinearity and dissipation to the formation of a shock wave can be characterized by a quantity similar to Reynolds number in hydrodynamics. In the nonlinear waves theory of classical media such a quantity is of the form [2]:

$\text{Re} = V\lambda\rho/2\pi b$, where $V$ is the velocity amplitude, $\lambda$ is the wave length, and the quantity $b$ is defined by viscosity and heat conductivity coefficients: $b = (4/3)\eta + \xi + \kappa(c_v^{-1} + c_p^{-1})$, $c_{v,p}$ are the heat capacities. We shall obtain the expression for an analogous quantity in superfluids basing on the following. Comparing the form of Burgers equation for classical media with that for superfluid helium one can see that the quantity $b/\rho$ in the classical case is replaced by the order of magnitude with the quantity $2\mu\rho_n/\rho_s$ in our case. Thus we may define the quantity Re as

$$\text{Re} = \frac{V_n \lambda \rho_s}{2\pi \, 2\mu \rho_n},$$

where $V_n$ is the wave amplitude in the temperature wave. In this formula $\mu$ characterizes dissipation and the value of $V_n$ is associated with nonlinearity.

We shall evaluate numerically the quantity Re since the solution of Eq. (8) depends on its magnitude. Let $T \approx 1{,}5^0$ K, $\omega \approx 2\pi \cdot 10^5 \, s^{-1}$, $V_n \approx 10$ cm/s (which corresponds to the wave intensity $\sim 10^{-4}$ W/cm$^2$); then, using the value of the wave amplitude damping coefficient, $\alpha \approx 10^{-1} cm^{-1}$ [5], we calculate that $\mu \approx 2 \cdot 10^{-3}$ cm$^2$/s, and as a result we obtain Re $\approx 10^2$. The analysis of Eq. (8) gives [2], that at Re>>1 a shock front is formed at a distance

$$x \geq \frac{\pi}{2} \frac{c_2^2}{\alpha_2 \, \omega V_n}. \qquad (9)$$

At shorter distances the squared velocity $v_{na}^2$ averaged over time does not depend on the coordinate and hence $F_2 = 0$.

The solution to Eq. (8) at distances where a steep shock front is formed is as follows (see [2])

$$v_{na} = \frac{V_n}{1+\varepsilon}\left[-\omega\tau + \pi \, \text{th}\left(\frac{\omega\tau}{\delta}\right)\right], \quad -\pi \leq \omega\tau \leq \pi, \qquad (10)$$

here $\varepsilon = \frac{\alpha_2}{c_2^2}\omega V_n x$, $\delta = \frac{1+\varepsilon}{\pi\alpha_2 \text{Re}}$. $\qquad (11)$

The quantity δ by the order of magnitude is the characteristic extent of the wave front divided by the wave length. We shall evaluate numerically this quantity. It is noted above that the solution is valid for $x$, defined by (9), that is for $\varepsilon \geq \pi/2$. For $\varepsilon \approx \dfrac{\pi}{2}$ and the above parameters (at $T \approx 1{,}5°$ K the quantity $\alpha_2 \approx 1$) we obtain $\delta \approx 10^{-2} \ll 1$. At higher $\delta$ values the wave front becomes smoother. Let us calculate the value of $\overline{v_{na}^2}$ (see (10)), which $x$-derivatives yield the force $F_2$, for the values of $\delta$ in the range from $\delta \ll 1$ to $\delta \sim 1$:

$$\overline{v_{na}^2} = \frac{V_n^2}{(1+\varepsilon)^2}\frac{1}{2\pi}\int_{-\pi}^{\pi}\left[-\omega\tau + \pi\,\text{th}(\omega\tau/\delta)\right]^2 d(\omega\tau). \qquad (12)$$

For $\delta \ll 1$ we can drop the second term in the integral (12), and we obtain (see also [2]):

$$\overline{v_{na}^2} = \frac{V_n^2}{(1+\varepsilon)^2}\frac{\pi^2}{3}.$$

For higher values of $\delta$ we shall calculate the integral (12) up to terms containing $\delta$ in higher degrees. Calculation yields the following result

$$\overline{v_{na}^2} = \frac{V_n^2}{(1+\varepsilon)^2}\left[\frac{\pi^2}{3} - \delta\pi + \frac{\delta^2\pi^2}{24} - \delta^2\,e^{-2\pi/\delta} + O(e^{-4\pi/\delta})\right]. \qquad (13)$$

One can see from (13), that at averaging $\overline{v_{na}^2}$ over time one may limit oneself with high accuracy with the terms to second order in $\delta$, that is with the first three terms. Taking into account that the force $F_2$ is the $x$-derivative of $\overline{v_{na}^2}$ one can easily see that only two first terms will remain in $F_2$, since the quantity $(1+\varepsilon)^2$, that depends on $x$ and enters the expression for $\delta^2$ (see (11)), is cancelled with the same quantity from the coefficient in front of the brackets. Further, numerical estimates show that the contribution of the second term in $F_2$ (6), which depends on viscosities, is significantly lower than that of the first term. This means that the force is caused mainly by energy dissipation at the shock wave front. As a result we find that the force $F_2$ that generates quasi-stationary normal flow in the presence of an intense temperature wave with high accuracy is defined by

$$F_2 = \frac{\rho_s}{\rho} \frac{\omega V_n^3}{c_2^2 (1+\varepsilon)^2} \left( \frac{2}{3} \frac{\pi^2 \alpha_2}{1+\varepsilon} - \frac{1}{\text{Re}} \right).$$

This is the force acting on a unit mass of the normal part. Let us estimate the possibility of experimentally observing the process. Let as above the second-sound intensity be of the order of magnitude ~ $10^{-4}$ *W/cm²* (the amplitude $V_n \approx 10$ *cm/s*), $\omega \approx 2\pi \cdot 10^5$ *s*$^{-1}$, $T \approx 1{,}5^0$ K (at this temperature the quantity $\alpha_2 \approx 1$, $\rho_s \approx \rho$), the interaction length $x \approx 1$ cm ($\varepsilon \approx \pi/2$). For these values the force $F_2 \approx 50$ *dyn/g*. We shall estimate what normal flow velocity could be induced by this force. Using Eq. (4) we can put approximately $\partial U_n / \partial t \approx F_2$. This yields that the flow velocity could reach the value (the initial value is zero) $U_n \approx F_2 \Delta t$, where $\Delta t$ is the pulse duration of the temperature wave. If $\Delta t$ is of the order of magnitude say $10^{-2}$ *s*, the velocity can be of the order of $5 \cdot 10^{-1}$ *cm/s* which is an experimentally observable value.

Thus we may conclude that in superfluid helium intense first-sound and temperature waves can induce stationary normal flow which might be specified as an entropy wind.

## References

1. L. K. Zarembo, V. A. Krasil'nikov. Introduction to nonlinear acoustics. Moscow: Nauka, 1966 (in Russian).
2. O. V. Rudenko, S. I. Soluyan. Theoretical foundations of nonlinear acoustics. Moscow: Nauka, 1975 (in Russian).
3. S. K. Nemirovskii. Nonlinear acoustics of superfluid helium // Sov. Phys. Usp. 1990. V. 33. № 6. P. 429 ─ 452.
4. L. D. Landau, E. M. Lifshitz. Fluid mechanics. Butterworth-Heinemann Ltd, 2000.
5. J. Wilks. The Properties of Liquid and Solid Helium, Oxford: Clarendon Press, 1967.